\documentclass[12pt]{article}
\usepackage{latexsym}
\usepackage{amssymb}
\usepackage{amsmath}
\usepackage{graphicx}
\usepackage{epsf}
\newcommand{\phis}{$\langle\phi^{2}\rangle$}
\newcommand{\texp}{$\langle T_{\mu \nu}\rangle$}

\begin{document}

\title{Gravitational Effects of Quantum Fields in the Interior of a 
Cylindrical Black Hole}
\author{\small Andrew DeBenedictis \footnote{e-mail:adebened@sfu.ca} 
\\ \it{\small Department of Physics} 
\\ \it{\small Simon Fraser University, Burnaby, British Columbia, Canada V5A 
1S6}} 
\date{October 18, 1998}
\maketitle

\begin{abstract}
The gravitational back-reaction is calculated for the conformally 
invariant scalar field within a black cosmic string interior with 
cosmological constant. Using the perturbed metric, the gravitational 
effects of the quantum field are calculated. It is found that the 
perturbations initially strengthen the singularity. This effect is 
similar to 
the case of spherical symmetry (without cosmological constant). This 
indicates that the behaviour 
of quantum effects may be universal and not dependent on 
the geometry of the spacetime nor the presence of a 
non-zero cosmological constant.
\end{abstract}

\vspace{3mm}
PACS number(s): 04.70.-s, 95.30.Sf \\

\section{Introduction}
\qquad The gravitational effects of quantum fields in black hole spacetimes 
has long been studied. Since Hawking's discovery that black holes radiate 
\cite{ref:hawkevap} much interesting work has been done in this area. 
Quantities of interest include the expectation value \phis, which describe 
vacuum polarization effects, and \texp, the expectation value of the 
stress-energy tensor of the field. This latter quantity may then be used 
in the Einstein Field equations:
\begin{equation}
R_{\mu\nu}-\frac{1}{2}Rg_{\mu\nu}+\Lambda g_{\mu\nu} = 8\pi\langle 
T_{\mu\nu} \rangle \label{eq:einst}
\end{equation}  
to determine the back-reaction of the field on the original spacetime. 
The effects of the back-reaction may also include the 
removal of singularities (\cite{ref:rem1}-\cite{ref:hiscock}). This is the 
main motivation for the work presented here and the answer would have 
consequences to many fundamental questions including the information loss 
problem.

\qquad Hiscock et.al \cite{ref:hiscock} have done an extensive 
study of 
these effects on the Schwarzschild interior and have found cases where 
curvature is initially slowed in the interior as well as cases where the 
curvature is initially strengthened (such as the case of the massless 
conformally coupled scalar field). They have also studied the effects on 
black hole anisotropy. It is interesting to ask whether or 
not the results are a product of the symmetry chosen or are general. It 
is also interesting to ask whether the presence of a cosmological 
constant will alter the situation. The study here attempts to address both 
issues by studying a black cosmic string which is asymptotically 
anti-deSitter. The field is in the Hartle-Hawking vacuum state 
\footnote{Due to the fact that the black string has positive specific 
heat, the Hartle-Hawking vacuum state is particularly applicable here. 
For a discussion of black string thermodynamic properties see
\cite{ref:brillthermo}, \cite{ref:mythermo} and \cite{ref:pecathermo}.} 
\cite{ref:harthawk} and the 
stress-energy tensor is found using the 
approximation of Page \cite{ref:pageappx} which is particularly useful 
here since the spacetime is an Einstein spacetime (in an Einstein
spacetime the relation $R_{\mu\nu}=\Lambda g_{\mu\nu}$ holds). Black 
string 
solutions are of relevance to cosmic strings. It has also been shown how 
such black holes may form by gravitational collapse \cite{ref:smithmann} 
\cite{ref:lemoscollapse}. This type of collapse has astrophysical 
relevance as the collapse of a finite spindle can behave as an infinite 
cylinder near its central region \cite{ref:spindle}.

\qquad It may be thought that, since no external observer can view the 
interior without falling into the black hole, that a study of the 
interior is not physically meaningful. However, as pointed out in 
\cite{ref:hiscock} black hole evaporation reveals more and more of the 
black hole interior as time progresses  and therefore the interior has 
relevance to exterior observers in this way. Also, the issue of whether or 
not spacetime singularities actually exist has been one of intense 
interest ever since Oppenheimer and Snyder's \cite{ref:opsnyd} original 
collapse calculation.

\section{Black String Spacetime}
\qquad The black hole studied here is the cylindrical black hole 
spacetime developed by Lemos and Zanchin \cite{lemos} and also studied by 
Kaloper \cite{kaloper}. If charge and angular momentum are not present 
the metric has the form
\begin{equation}
ds^{2}=-(\alpha^{2}\rho^{2}-\frac{4M}{\alpha\rho}) dt^{2} + 
\frac{d\rho^{2}}{(\alpha^{2}\rho^{2}-\frac{4M}{\alpha\rho})} + \rho^{2} 
\,d\varphi^{2} + \alpha^{2}\rho^{2}\, dz^{2}. \label{eq:metric}
\end{equation}
where $M$ is the mass per unit length, $\alpha^{2}=-\frac{1}{3}\Lambda$ and 
the coordinates take on the following ranges:
\begin{eqnarray}
-\infty < t < \infty , \nonumber \\
0 \leq \rho < \infty , \nonumber \\
0\leq \varphi <2\pi , \nonumber \\
-\infty < z < \infty. \nonumber
\end{eqnarray}
An event horizon exists at $\rho=\rho_{H}\equiv 
\frac{(4M)^{1/3}}{\alpha}$ and the cosmological constant 
(which is negative and necessary for cylindrical black hole solutions), 
$\Lambda$, dominates in the limit $\rho\rightarrow\infty$ giving the 
spacetime its asymptotically anti-deSitter behaviour.

\qquad The apparently singular behaviour of the spacetime at 
$\rho=\rho_{H}$ is a coordinate effect and not a true singularity. On 
calculating the Kretschmann scalar one obtains
\begin{equation}
K\equiv 
R_{\delta\lambda\mu\nu}R^{\delta\lambda\mu\nu} = 
24\alpha^4\left(1+\frac{8M^{2}}{\alpha^{6}\rho^{6}}\right) \label{eq:origK}
\end{equation}
from which it can be seen that the only true singularity is a polynomial 
singularity at $\rho=0$. Thus, this solution violates the hoop conjecture 
but not the cosmic censorship conjecture. The hoop conjecture is 
therefore not valid in spacetimes with a cosmological constant.

As calculations will be extended to the interior, it is convenient to 
re-write the metric using the following coordinate redefinitions:
\begin{eqnarray}
t \rightarrow R , \nonumber \\
\rho \rightarrow T. \nonumber
\end{eqnarray}
Where $T$ is timelike in the interior and $R$ is spacelike. The ``interior" 
metric now has the form \begin{equation}
ds^{2}_{interior}=-\frac{dT^{2}}{\left(\frac{4M}{\alpha T} - 
\alpha^{2}T^{2}\right)} + \left(\frac{4M}{\alpha T} 
-\alpha^{2}T^{2}\right) dR^{2}+ T^2\,d\varphi^{2} + \alpha^{2}T^{2} \, 
dz^{2} \label{eq:intmet}
\end{equation}
where the interior region corresponds to $0 \leq T \leq \rho_{H}$.

\section{Stress-Energy Tensor}
\qquad In this section the stress energy tensor is calculated which will 
eventually be used in (\ref{eq:einst}) to calculate back-reaction 
effects. The expectation value of stress-energy tensors have been 
calculated in exterior Schwarzschild spacetime by Howard and Candelas 
\cite{ref:handctmn} and Page \cite{ref:pageappx} as well as by Anderson 
et. al \cite{ref:ahl} who
studied the stability in the extreme Reissner-Nordstr\"{o}m black hole. 
Anderson, Hiscock and Samuel \cite{ref:ahs} have 
developed an approximation for both massive and massless fields in arbitrary 
spherically symmetric spacetimes and have used this approximation to 
calculate \texp\ in the exterior Reissner-Nordstr\"{o}m geometry. The 
Kerr and Kerr-Newman spacetimes have also been studied in \cite{ref:frolov}, 
\cite{ref:frovandthorne} and \cite{ref:frovandzel}. Quantum effects in 
lower dimensional black hole exteriors may be found
in \cite{ref:btz1}-\cite{ref:btz10}. For a calculation of 
\phis\ in the spacetime studied here see \cite{ref:myphi}. 

\qquad Various works on back-reaction effects of quantum fields have also 
been produced. Hiscock and Weems \cite{ref:handw}, Bardeen \cite{ref:B1}, 
Balbinot \cite{ref:B2} and York \cite{ref:york} have studied 
effects in Schwazschild and Reissner-Nordstr\"{o}m exteriors. Few 
calculations, however, have been performed on the interiors of black 
holes. One such study has been done by Hiscock, Larson 
and Anderson \cite{ref:hiscock} where they have extended their analysis to 
the Schwarzschild interior and calculated back-reaction effects on curvature 
invariants.

\subsection{Stress-Energy Tensor for the Conformally Coupled Scalar Field}

\qquad The calculation of the stress energy tensor will be done using 
the Eucldeanized metric. This is obtained by making the transformation 
($t\rightarrow -i\tau$) in (\ref{eq:metric}) giving the metric positive 
definite signature so that 
\begin{equation}
ds^{2}_{Euclidean}=(\alpha^{2}\rho^{2}-\frac{4M}{\alpha\rho}) d\tau^{2} +
\frac{d\rho^{2}}{(\alpha^{2}\rho^{2}-\frac{4M}{\alpha\rho})} + \rho^{2}
\,d\varphi^{2} + \alpha^{2}\rho^{2}\, dz^{2}. \label{eq:eucmetric}
\end{equation}
To calculate \texp\ exactly is an extremely 
difficult task 
which normally involves acting on \phis\ with a complicated differential 
operator. It is useful therefore to use an approximation which will give 
an analytic result from which information on back-reaction effects may be 
calculated. The approximation used here is the approximation of Page for 
thermal stress-energy tensors in static spacetimes\cite{ref:pageappx}. This 
approximation is especially good if the spacetime under consideration is 
an Einstein spacetime such as the one considered here and contains no 
ambiguities in the case of scalar fields. The 
Bekenstein-Parker \cite{ref:bekpark} Gaussian path integral approximation 
is utilized from which the thermal propagator is constructed. This 
construction is done in an (Euclideanized) ultrastatic spacetime 
($g_{00}=k,\,\, k$ 
is a constant chosen to be $1$ in this work) which is related to the 
physical spacetime by 
\begin{equation}
g_{\mu\nu}=|g_{00\,\,(p)}|^{-1}g_{\mu\nu\,\,(p)}.
\end{equation}
The subscript $p$ will be used to indicate quantities calculated using the 
physical metric (all other tensors in this section are obtained using the 
ultrastatic metric).
This approximation gives, for the stress-energy tensor in the physical 
spacetime:
\begin{eqnarray}
T^{\mu}_{\nu\,\, (p)}=|g_{00\,\, (p)}|^{-2} \lbrace T^{\mu}_{\nu} + 
[8\lambda |g_{00\,\,(p)}|^{-1}(|g_{00\,\,(p)}|^{1/2})_{;\alpha} 
(|g_{00\,\,(p)}|^{1/2})^{;\beta} \nonumber \\ 
-4(\lambda +\beta) R^{\beta}_{\alpha}]C^{\alpha\mu}_{\beta\nu}+ 
2\beta [H^{\mu}_{\nu}+3\alpha^{4}|g_{00\,\,(p)}|^{2}\delta^{\mu}_{\nu}]+ 
\frac{1}{6}\gamma I^{\mu}_{\nu}\rbrace , \label{eq:tmn1}
\end{eqnarray}
where $C^{\alpha\mu}_{\beta\nu}$ is the Weyl tensor 
and the coefficients $\lambda,\,\beta$ and $\gamma$ are given as follows: 
\begin{equation} \lambda=\frac{12h(0)}{2^{9}45\pi^{2}},
\,\,\beta=\frac{-4h(0)}{2^{9}45\pi^{2}},
\,\,\gamma=\frac{8h(0)}{2^{9}45\pi^{2}}. \end{equation}
The number of helicity states, $h(0)$, simply counts the number of scalar 
fields present.
$T^{\mu}_{\nu}$ is the stress-energy tensor in the ultrastatic metric, 
\begin{equation}
T^{\mu}_{\nu}=\frac{\pi^{2}}{90}T^{4}(\delta^{\mu}_{\nu}- 
4\delta^{\mu}_{0}\delta^{0}_{\nu}),
\end{equation}
were $T$ is the temperature of the black string which can be found by 
demanding 
that the Euclidean extension of (\ref{eq:metric}) be regular on the 
horizon; 
\begin{equation}
T=\frac{3\alpha}{4\pi}(4M)^{1/3}. \label{eq:temp}
\end{equation}
The quantities $H^{\mu}_{\nu}$ and $I^{\mu}_{\nu}$ are given by:
\begin{eqnarray}
H^{\mu}_{\nu}=-R^{\alpha\mu}R_{\alpha\nu}+ \frac{2}{3}RR^{\mu}_{\nu} + 
(\frac{1}{2}R^{\alpha}_{\beta}R^{\beta}_{\alpha}- 
\frac{1}{4}R^{2})\delta^{\mu}_{\nu}, \nonumber \\
I^{\mu}_{\nu}=2R^{;\mu}_{;\nu}-2RR^{\mu}_{\nu}+(\frac{1}{2}R^2- 
2R^{;\alpha}_{;\alpha})\delta^{\mu}_{\nu}. \label{eq:handI}
\end{eqnarray}

\qquad The calculation of \texp\ is carried out on the exterior of the 
black hole. However, since the result is finite at the horizon, it is 
easily extended to the interior where the field equations will be solved. 
For the spacetime considered here, the stress-energy tensor is calculated 
to be
\begin{eqnarray}
T^{\mu}_{\nu\,\,(p)}=-\lbrace 
1920\pi^{2}\epsilon 
[\alpha^{4}\rho^{6}(\alpha^{3}\rho^{3}-4M)^{2}]\rbrace^{-1} 
[-27\,2^{2/3}\,M^{4/3}\alpha^{8}\rho^{8}(\delta^{\mu}_{\nu}- 
4\delta^{\mu}_{0}\delta^{0}_{\nu}) \nonumber \\ 
-16M\alpha^{9}\rho^{9} 
(3\delta^{\mu}_{0}\delta^{0}_{\nu}+\delta^{\mu}_{1}\delta^{1}_{\nu})
-128\alpha^{3}\rho^{3}M^{3}(\delta^{\mu}_{\nu} -12\delta^{\mu}_{0} 
\delta^{0}_{\nu}) \nonumber \\ 
+192M^{4}(\delta^{\mu}_{\nu}- 12\delta^{\mu}_{0} 
\delta^{0}_{\nu}- \frac{8}{3}\delta^{\mu}_{1}\delta^{1}_{\nu}) \nonumber \\
+96M^{2}\alpha^{6}\rho^{6}(\delta^{\mu}_{\nu} -5\delta^{\mu}_{0} 
\delta^{0}_{\nu} +\delta^{\mu}_{1}\delta^{1}_{\nu})
+2\alpha^{12}\rho^{12}\delta^{\mu}_{\nu}], \label{eq:setensor}
\end{eqnarray}
where $\epsilon= \hbar\alpha^{2}$.
This function remains unchanged when analytically continued to the 
Lorentzian sector by the transformation $\tau \rightarrow it$ and 
has trace consistent with anomaly calculations.
Far from the black string, (\ref{eq:setensor}) takes on its pure 
anti-deSitter value of 
$-\frac{\alpha^{4}}{960\pi^{2}}\delta^{\mu}_{\nu}$ \cite{ref:pbo} whereas 
at the horizon (\ref{eq:setensor}) is also well defined and given  by 
\begin{equation} 
T^{\mu}_{\nu\,\,(p)}(\rho_H)= 
\frac{\alpha^{4}}{\pi^{2}}\left(\begin{array}{cccc}
\frac{1}{640}& 0 & 0 & 0 \\
0 & \frac{1}{640}& 0 & 0 \\
0 & 0 & -\frac{1}{640} & 0 \\
0 & 0 & 0 & -\frac{1}{640} \end{array} \right). \label{eq:tmnhorz}
\end{equation}
Inspection of (\ref{eq:tmnhorz}) immediately shows that the weak energy 
condition (WEC) is violated. The qualitative behaviour of the energy density 
($\varepsilon=-T^{0}_{0}$) is shown in figure 1 where it can be seen that 
the WEC is violated throughout the interior of the black hole ($\rho< 
\sim 1.6$). However, it is unknown how relevant the classical energy 
conditions are in the case of quantum matter where violations are common 
(for example in the case of the Casimir effect) and are in fact required 
for a self consistent picture of Hawking evaporation.

\begin{figure}[ht]
\label{fig1}
\includegraphics[bb=70 286 507 720, width=0.8\textwidth,clip]{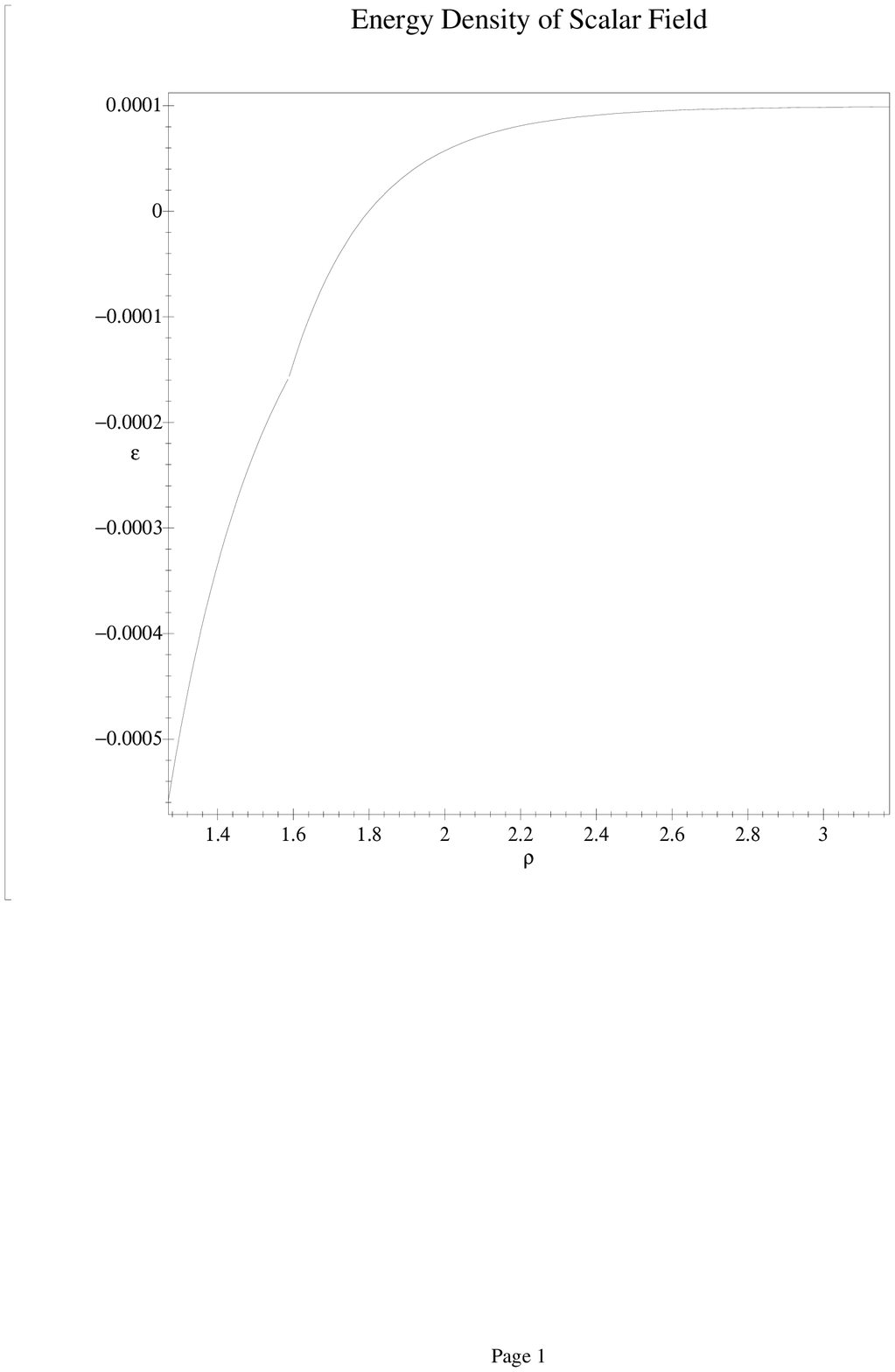}
\caption{{\small Energy density of the quantum scalar field in the 
cylindrical 
black hole spacetime. Weak energy condition violation can be seen 
throughout the interior ($\rho<\sim 1.6$) and part of the exterior. The 
interior energy density is given by $-T^{T}_{T}$ whereas on the exterior 
it is given by $-T^{t}_{t}$.}}
\end{figure}  
\section{Gravitational Back-Reaction}
\qquad In this section the gravitational effects of the quantum field on 
the background spacetime will be calculated using the perturbed metric
\begin{eqnarray}
ds^2=-\frac{dT^2}{\frac{4M}{\alpha T}-\alpha^2 
T^2}\left(1+\epsilon\eta (T)\right) &+& \left(\frac{4M}{\alpha T}-\alpha^2 
T^2\right)\left(1+\epsilon\sigma (T)\right)\,dR^{2} \nonumber \\
&+& T^2\,d\varphi^{2}+\alpha^{2}T^{2}\left(1+\epsilon\psi 
(T)\right)\,dz^{2}. \label{eq:pertmet}
\end{eqnarray}
The functions $\eta(T)$, $\sigma(T)$ and $\psi(T)$ are to be solved for and
the coupling constant, $\epsilon$ is assumed to 
be small. The Einstein field equations, to first order in $\epsilon$ yield
\begin{subequations}
\begin{align}
& \sigma (T)'(\alpha^{3}T^{3}-4M)+\psi (T)'(\alpha^{3} T^{3}+M) = 
\left(\frac{8\pi 
\langle T^{T}_{T} \rangle}{\epsilon} +3\alpha^{2} \eta (T)\right)\alpha 
T^{2}, \label{eq:E1} \\
& \frac{d}{dT}\left[\eta (T)(4M-\alpha^{3} T^{3}) + \psi 
(T)'\left(\frac{\alpha^{3}T^{4}}{2}-2TM\right)\right]-3M\psi (T)' 
\nonumber \\
& = \frac{8\pi \langle T^{R}_{R} \rangle \alpha T^{2}}{\epsilon} , 
\label{eq:E2} \\ 
& \frac{d}{dT}\left[(\sigma 
(T)'+\psi (t)')\left(\frac{\alpha^{3}T^{4}}{2}-2TM\right) 
\right]+3M\sigma (T)'+\eta (T)'(M-\alpha^{3}T^{3}) \nonumber \\ 
& = \left(\frac{ 8\pi\langle T^{\varphi}_{\varphi}\rangle}{\epsilon} 
+ 3 \alpha^{2}\eta (T)\right) \alpha T^{2}, 
\label{eq:E3} \\
& \frac{d}{dT}\left[\sigma 
(T)'\left(\frac{\alpha^{3}T^{4}}{2}-2TM\right)\right] +3M\sigma (T)' + 
\eta (T)'(M-\alpha^{3}T^{3}) \nonumber \\ 
&= \left(\frac{ 8\pi \langle T^{z}_{z} \rangle}{\epsilon} + 3 
\alpha^{2}\eta (T)\right)\alpha T^2, \label{eq:E4}
\end{align}
\end{subequations}
where primes denote ordinary differentiation with respect to $T$.

\qquad Since solving (\ref{eq:E1}-\ref{eq:E4}) using (\ref{eq:setensor}) 
and calculating the resulting relevant quantities such as the Riemann 
curvature tensor and Kretschmann scalar would be an enourmous task, some 
simplifying assumptions are first made. It is noted from 
(\ref{eq:setensor}) that equations (\ref{eq:E3}) and (\ref{eq:E4}) must 
be equal. It is therefore assumed that the function $\psi(T)$ is equal to 
a constant and therefore does not appear in the field equations. 
Secondly, the stress energy tensor will be approximated by its value near 
the event horizon. This should not introduce too large of an error in the 
calculations as the perturbative scheme here is only valid in regions 
where the spacetime curvature is not large (such as near the horizon in 
the small $\alpha$ limit).

\qquad Utilizing the above, the following solutions are obtained for the 
perturbations:
\begin{eqnarray}
\eta (T)&=&\frac{1}{\pi(4M-\alpha^{3}T^{3})} \left[ \frac{13}{240} 
\alpha^{3}T^{3}-\frac{3}{160}\alpha^{4}T^{4}M^{-1/3}-\frac{M}{15}\right] 
\label{eq:eta} \\
\sigma(T)&=&-\frac{1}{\pi(4M-\alpha^{3}T^{3})} \left[\frac{13}{240}
\alpha^{3}T^{3}-\frac{3}{160}\alpha^{4}T^{4}M^{-1/3}- \frac{M}{15}\right] 
\nonumber \\ 
&+& \frac{1}{30}\frac{\alpha T 2^{1/3}}{\pi 
M^{1/3}} 
+ \frac{1}{45\pi}\ln\left(\frac{(4M)^{1/3}-\alpha T}{4M-\alpha^3 T^3}\right) 
\nonumber \\ 
&-& \frac{1}{90\pi}\ln (\alpha^{2}T^{2}+\alpha
T(4M)^{1/3}+(4M)^{2/3}) \nonumber \\
&-&\frac{\sqrt{3}}{45\pi}\arctan\left[\frac{1}{\sqrt{3}}\left( \frac{\alpha T
2^{1/3}}{M^{1/3}}+1\right)\right]+k_{0}. 
\label{eq:sigma}
\end{eqnarray}

The integration constant $k_{0}$ may be left arbitrary as it does not 
enter subsequent calculations. Both solutions are well behaved in the 
domain of validity as $\eta(\rho_H)=\frac{1}{240\pi}$ and 
$\sigma(\rho_H)=\frac{153-72\ln(3)-96\ln(2)+48\ln(M)+16\sqrt{3}\pi}{2160\pi}$.

\qquad Attention is now turned to the effects of the quantum perturbation 
on the black hole spacetime. It has long been thought that quantum 
effects may remove the singular behaviour of physical spacetimes. 
Although the perturbative scheme can not determine  whether or not the 
actual singularity is removed, it can give information on the growth of 
curvature scalars on the interior spacetime. Using the perturbed metric, 
the Kretschmann scalar can now be written as:
\begin{equation}
K=K_{orig} + \epsilon\delta K,
\end{equation}
where $K_{orig}$ is the unperturbed value given by (\ref{eq:origK}) and 
$\delta K$ is the first order correction term.
Whether or not curvature is strengthened depends on the sign of  $\delta
K$. If it is positive, the initial curvature growth is 
strengthened. If it is negative, it is weakened. The analysis will be 
limited to the region near the horizon.

\qquad The correction term is calculated to be
\begin{eqnarray}
\delta K &=&\frac{-1}{15}\left[25\,\alpha^{17}T^{17} 
2^{2/3}+54\,\alpha^{16}T^{16}(2M)^{1/3}+116\,\alpha^{15}T^{15}M^{2/3} 
\right. \nonumber \\
&-&286\,\alpha^{14}T^{14} 2^{2/3}M- 624\,\alpha^{13}T^{13}2^{1/3}M^{4/3}- 
1352\,\alpha^{12}T^{12}M^{5/3}\nonumber \\ 
&+& 1104\,\alpha^{11}T^{11}2^{2/3}M^2+ 
2256\,\alpha^{10}T^{10}2^{1/3}M^{7/3}  
4608\,\alpha^{9}T^{9}M^{8/3} \nonumber \\ 
&-& 2656\,\alpha^{8}T^{8} 2^{2/3}M^{3}- 
3456\,\alpha^{7}T^{7}2^{1/3}M^{10/3}- 
3200\,\alpha^{6}T^{6}M^{11/3} \nonumber \\
&+& 9472\,\alpha^{5}T^{5}2^{2/3}M^{4}+ 9984\,\alpha^{4}T^{4}2^{1/3}M^{13/3} 
+2048\,\alpha^{3}T^{3}M^{14/3}\nonumber \\ 
&-& \left.18432\, \alpha^{2}T^{2}2^{2/3}M^{5}- 
24576\,\alpha T 2^{1/3}M^{16/3}-24576M^{17/3}\right] \nonumber \\
&\times&\left[\alpha^{2}T^{6}M^{1/3}\pi (\alpha^{2}T^{2}+\alpha T (4M)^{1/3} 
+ (4M)^{2/3})^{2} \right. \nonumber \\
&\times& \left. (4M^{2/3}+\alpha^{2} T^{2} 2^{2/3} + 2\alpha T 
(2M)^{1/3})^{2} (\alpha T- (4M)^{1/3})^{2}\right]^{-1}  
\end{eqnarray}
Although this result is slightly complicated, a few relevant properties 
may be obtained. The value of 
$\delta K$ near the horizon behaves as 
\begin{equation} \delta K \approx \frac{3}{10}\frac{\alpha^{4}}{\pi}-
2\frac{\alpha^{5}}{\pi}\left(\frac{2}{M}\right)^{1/3} \left(T-
\frac{(4M)^{1/3}}{\alpha}\right), \label{eq:expdK}
\end{equation}
so that the limiting value at $T = \rho_{H}$ is given by
$\delta K(\rho_{H})= \frac{3}{10}\frac{\alpha^{4}}{\pi}$.
from which it can be seen that curvature growth is {\em strengthened} at the 
horizon. The function (\ref{eq:expdK}) increases as one passes 
through the 
horizon towards the singularity and the perturbation eventually becomes 
positive. Near the singularity the curvature diverges as $1/T^{16}$  
although in this regime the approximation breaks down and the expression 
has no physical meaning. 

\qquad For the case of Schwarzschild geometry the curvature perturbation 
diverges as $1/T^{9}$ near the singularity for massless conformally coupled 
scalars \cite{ref:hiscock} and in the regime where the perturbation is 
valid, curvature invariants are always strengthened. At the event 
horizon 
of a Schwarzschild black hole, for example, the perturbation is $\delta 
K= \frac{1965}{2M^{4} 45\times 2^{13}\pi}$.

\section{Conclusion}
The stress energy tensor for a conformally coupled quantum scalar field 
has been calculated in the black string spacetime and it is found that, 
as is common with quantum 
fields in curved spacetime, there exist regions where the weak energy 
condition is violated. The 
violation occurs on the interior and near the horizon on the exterior of 
the black hole. From the stress energy tensor, the back-reaction has been 
calculated in the form of the perturbed metric and Kretschmann scalar. 
Similar to the case of spherical symmetry without cosmological constant, it 
is found that curvature is strengthened on the interior indicating 
that quantum effects may be geometry and cosmological constant independent.

\section{Acknowledgments}
The author would like to thank Dr. K.S. Viswanathan for helpful advice 
during the production of this work. Discussions with Dr. A. Das on 
relativity were also helpful.

\newpage
\bibliographystyle{unsrt}

\begin{thebibliography}{10}

\bibitem{ref:hawkevap}
Hawking~S~W
\newblock 1975 {\em Commun. Math. Phys.} {\bf 43} 199

\bibitem{ref:rem1}
Parker~L and Fulling~S
\newblock 1973 {\em Phys. Rev. D}{\bf 7} 2357 

\bibitem{rem2}
Starobinsky~A~A
\newblock 1980 {\em Phys. Lett}{\bf 91B} 99

\bibitem{rem3}
Fischetti~M~V, Hartle~J~B and Hu~B~L
\newblock 1979 {\em Phys. Rev. D}{\bf 20} 1757

\bibitem{rem4}
Anderson~P
\newblock 1983 {\em Phys. Rev. D}{\bf 28} 271

\bibitem{rem5}
Azuma~T and Wada~S
\newblock 1986 {\em Prog. Theor. Phys.}{\bf 75} 845

\bibitem{ref:hiscock}
Hiscock~W~A, Larson~S~L and Anderson~P~R
\newblock 1997 {\em Phys. Rev. D}{\bf 56} 3571

\bibitem{ref:brillthermo}
Brill~D~R Louko~J and Peldan~P
\newblock 1997 {\em Phys. Rev. D}{\bf 56} 3600

\bibitem{ref:mythermo}
DeBenedictis~A
\newblock gr-qc/9809025 ~1998

\bibitem{ref:pecathermo}
Peca~C~S and Lemos~J~P~S
\newblock gr-qc/9809029 ~1998

\bibitem{ref:harthawk}
Hartle~J~B and Hawking~S~W
\newblock 1976 {\em Phys. Rev. D}{\bf 13} 2188

\bibitem{ref:pageappx}
Page~D~N
\newblock 1982 {\em Phys. Rev. D}{\bf 25} 1499

\bibitem{ref:smithmann}
Smith~W~L and Mann~R~B
\newblock 1997 {\em Phys. Rev. D}{\bf 56} 4942

\bibitem{ref:lemoscollapse}
Lemos~J~P~S
\newblock 1998 {\em Phys. Rev. D}{\bf 57} 4600

\bibitem{ref:spindle}
Shapiro~S~L and Teukolsky~S~A
\newblock 1991 {\em Phys. Rev. Lett.}{\bf 66} 944

\bibitem{ref:opsnyd}
Oppenheimer~J~R and Snyder~H
\newblock 1939 {\em Physical Review}{\bf ~56} 455

\bibitem{lemos}
Lemos~J~P~S and Zanchin~V~T
\newblock 1996 {\em Phys. Rev. D}{\bf 54} 3840

\bibitem{kaloper}
Kaloper~N
\newblock 1993 {\em Phys. Rev. D}{\bf 48} 4658

\bibitem{ref:handctmn}
Howard~K~W and Candelas~P
\newblock 1984 {\em Phys. Rev. Lett.}{\bf 53} 403

\bibitem{ref:ahl}
Anderson~P~R, Hiscock~W~A and Loranz~D~J
\newblock 1995 {\em Phys. Rev. Lett.}{\bf 74} 4365

\bibitem{ref:ahs}
Anderson~P~R, Hiscock~W~A and Samuel~D~A
\newblock 1995 {\em Phys. Rev. D}{\bf 51} 4337

\bibitem{ref:frolov}
Frolov~V~P
\newblock 1982 {\em Phys. Rev. D}{\bf 21} 2185

\bibitem{ref:frovandthorne}
Frolov~V~P and Thorne ~K~S
\newblock 1989 {\em Phys. Rev. D}{\bf 39} 2125

\bibitem{ref:frovandzel}
Zel'nikov~A~I and Frolov~V~P
\newblock 1987 Proceedings of the Lebedev Physics Institute of the 
Academy of Sciences of the USSR
\newblock {\bf 169} Nova Science, New York 171

\bibitem{ref:btz1}
Banados~M, Teitelboim~C and Zanelli~J
\newblock 1992 {\em Phys. Rev. Lett.}{\bf 69} 1849

\bibitem{btz2}
Lifschytz~G and Oritz~M
\newblock 1994 {\em Phys. Rev. D}{\bf 49} 1929

\bibitem{btz3}
Shiraishi~K and Maki~T
\newblock 1994 {\em Class. Quant. Grav.}{\bf 11} 695

\bibitem{btz4}
Shiraishi~K and Maki~T
\newblock 1994 {\em Phys. Rev. D}{\bf 49} 5286

\bibitem{btz5}
Shiraishi~K and Maki~T
\newblock 1994 {\em Class. Quant. Grav.}{\bf 11} 1687

\bibitem{btz6}
Steif~A
\newblock 1994 {\em Phys. Rev. D}{\bf 49} 585

\bibitem{btz7}
Brown~J~D, Creighton~J and Mann~R~B
\newblock 1994 {\em Phys. Rev. D}{\bf 50} 6394

\bibitem{btz8}
Reznik~B
\newblock 1995 {\em Phys. Rev. D}{\bf 51} 1728

\bibitem{btz9}
Mann~R~B and Solodukhin~S~N
\newblock 1997 {\em Phys. Rev. D}{\bf 55} 3622

\bibitem{ref:btz10}
Oda~I
\newblock 1997 {\em Phys. Lett.}{\bf B409} 88

\bibitem{ref:myphi}
DeBenedictis~A
\newblock gr-qc/9804032

\bibitem{ref:handw}
Hiscock~W~A and Weems~L
\newblock 1990 {\em Phys. Rev. D}{\bf 41} 1142

\bibitem{ref:B1}
Bardeen~J~M
\newblock 1981 {\em Phys. Rev. Lett.}{\bf 46} 382

\bibitem{ref:B2}
Balbinot~R
\newblock 1984 {\em Class. Quant. Grav.}{\bf 1} 573

\bibitem{ref:york}
York~J~W Jr.
\newblock 1985 {\em Phys. Rev. D}{\bf 31} 775

\bibitem{ref:bekpark}
Bekenstein~J~D and Parker~L
\newblock 1981 {\em Phys. Rev. D}{\bf 23} 2850

\bibitem{ref:pbo}
Brown~M~R, Ottewill~A~C and Page~D~N
\newblock 1986 {\em Phys. Rev. D}{\bf 33} 2840

\end{thebibliography}

\end{document}